# A zero point energy explanation of a peak in liquid helium's dynamic structure factor

*Max Chaves*

*Escuela de Fisica*

*Universidad de Costa Rica*

*Ciudad Universitaria Rodrigo Facio*

*San Jose, Costa Rica*

**Recent high resolution experiments show a strong peak at $Q = 1.9 \text{Å}^{-1}$ in liquid helium's dynamic structure factor that exhibits a singular dependence on temperature. The theoretical situation is briefly reviewed, and the comment is made that the simplest possible explanation is that at the lambda point a macroscopic number of molecules actually begin having a wavenumber $k = Q \approx 1.9 \text{Å}^{-1}$ . It is pointed out that the intermolecular separation at the lambda point is the right one, considering the size of helium molecules, to *almost* form around each molecule a *three-dimensional* box made of other molecules. A trapped molecule has, according to the uncertainty principle, $k \simeq 1.9 \text{Å}^{-1}$ . An experiment with liquid helium is suggested that can help clarify the nature of the peak.**

PACS numbers 67.40  25.40.D  03.75.Fi

# 1. A SINGULAR EXPERIMENTAL RESULT

An interesting experimental result in neutron scattering by liquid helium ($^4$He) been observed in the last few years. It concerns high resolution data pertaining the dynamic structure factor $S(\boldsymbol{Q}, \omega)$, one of the fundamental quantities involved in the analysis of neutron scattering experiments. The arguments of the dynamic structure factor are the wavevector $\boldsymbol{Q}$ and the angular frequency $\omega$, which, multiplied by Planck's constant, are the momentum and energy transferred into liquid helium by the scattered neutron:

$$\hbar\boldsymbol{Q} = \boldsymbol{p}_f - \boldsymbol{p}_i,$$
$$\hbar\omega = E_f - E_i.$$

(1)

The $i$ and $f$ subindices refer respectively to the initial and final values of the momentum and energy of the neutron. $S(\boldsymbol{Q}, \omega)$ is defined in terms of the phase space number density $\hat{\rho}(\boldsymbol{Q}, t)$ by

$$S(\boldsymbol{Q}, \omega) = \frac{1}{2\pi N} \int_{-\infty}^{\infty} dt \, e^{i\omega t} \langle \hat{\rho}(\boldsymbol{Q}, t) \hat{\rho}^*(\boldsymbol{Q}, t) \rangle \,,$$

(2)

where $N$ is the number of molecules. Furthermore it can be shown that, in the first Born approximation,

$$\frac{d^2\sigma}{d\Omega dE} \propto S\left( \boldsymbol{Q}, \omega \right).$$

(3)

So what is done is to use neutron scattering to study the structure form factor, which in turn gives information about the excitations of liquid helium.

Recent high resolution experiments have allowed the study in detail the temperature dependence of peaks of $S(\boldsymbol{Q}, \omega)$ at $Q = 0.4 \text{Å}^{-1}$ and $Q = 1.9 \text{Å}^{-1}$. Their existence has been known [1] for some time, and what is new [2,3] is the detail with which we now can know their dependence on temperature. For a long time the first peak has been associated with phonons. It is basically temperature independent. The second peak begins forming only after the temperature has dropped down to $T_\lambda$, the lambda point; it is not there for $T > T_\lambda$. As the temperature drops even further it keeps increasing until at absolute zero it involves about $10\%$ of the liquid.

It is usually assumed [4] that in liquid helium a condensate $n_0$ begins forming at the lambda point, so the experimental result mentioned, where the peak begins forming at wavenumber $Q = 1.9 \text{Å}^{-1}$ and not at $Q = 0 \text{Å}^{-1}$ like expected, demands our attention. One effort in trying to understand it is the following scenario, [5] that I will call the pole hybridization picture (PHP). In it the dispersion curve is assumed to be due to two different branches, both independent of the existence of the condensate. The 'phonon' branch results in the sector of the dispersion curve of the same name (which corresponds to vanishing values of $Q$), and is understood as due to zero sound. (This concept was extended to Bose particles by Pines. [6]) It appears as a pole of the dynamic structure factor. There is also in the PHP a 'roton' branch of single particle origin that results in the sector of the dispersion curve of the same name. Rotons are a pole of the single particle Green function $G(\boldsymbol{Q}, \omega)$, that, through hybridization due to the condensate $n_0$, appear as a pole of $S(\boldsymbol{Q}, \omega)$. When the condensate vanishes at $T > T_\lambda$, hybridization ceases to exist, likewise the roton peak of the dynamic structure function. Nevertheless, according to PHP, the single particle branch is still there. In this picture, the roton peak does not exist for $T > T_\lambda$ because the condensate $n_0$ does not either. [7]

The PHP has been called into question, and instead the modified classical picture [8] (MCP) has been suggested. It is shown in the MCP reference on the basis of quantum field theory arguments that:

- there is a qualitative difference between the excitations above and below $T_\lambda$;
- the phonon and roton branches have a unified character and are both of single particle origin.

In the MCP, hybridization is still accepted, but it is emphasized that it implies the suppression of the zero sound branch for $T < T_\lambda$. This leaves only the single particle sound branch for this low-temperature regime. The theoretical arguments given in the MCP reference against the PHP seem fairly strong and correct. The main conclusion we derive from the MCP is that the PHP is not a self-consistent explanation of the observed experimental results. On the other hand, we are left with a feeling of dissatisfaction as to what is the actual physics behind the temperature dependence of the two peaks.

Let us consider the situation under a more general light. In the experiments we see the formation of a strong peak at $Q = 1.9 \text{Å}^{-1}$ when the temperature drops down to $T = T_\lambda$. Since we are assuming that a condensate $n_0$ begins to form at precisely this temperature, we invoke the hybridization process to explain how a condensate forming at $Q = 0 \text{Å}^{-1}$ shows up as a peak forming at $Q = 1.9 \text{Å}^{-1}$. But there is another simpler explanation for what is being seen: the $1.9 \text{Å}^{-1}$ peak in $S(\boldsymbol{Q}, \omega)$ forms because at $T = T_\lambda$ more and more excitations begin to have precisely this wavenumber, that is, it appears to form at this wavenumber because it is actually doing so, and it is interesting that there is also a simple physical mechanism that produces precisely this effect.

## 2. A PHYSICAL MECHANISM THAT PRODUCES A PEAK AT $Q = 1.9 \text{Å}^{-1}$

It was suggested [9] some time ago that a nonzero root-mean-square superfluid velocity due to quantum fluctuations exists in a superfluid at rest. There is no net liquid transport because the current vector is always switching directions. It also has been proposed [10] on the basis of theoretical considerations of the one-particle spectrum, that there could be a Bose-Einstein condensate at a nonzero wavenumber. Later it was also pointed out [11] that the experimentally successful Hyland-Rowlands-Cummings formula [12] could be explained assuming such a condensate. This is convenient since the original derivation has been shown to be incorrect. [13] Here we show, in agreement with these ideas, that it is likely that a macroscopic number of helium molecules acquire a rms wavenumber $k = Q \approx 1.9 \text{Å}^{-1}$ at the lambda point, so that the peak seen at this momentum transfer $Q$ in $S(\boldsymbol{Q}, \omega)$ is to be expected.

It is well-known that a liquid helium molecule has a zero point energy $E_0$ that is in the order of its average kinetic energy $E$. The zero point energy is the minimum kinetic energy imposed on a molecule by its inherent wave-like nature and the size of the volume it is enclosed in. For most materials $E_0 \ll E$, but in helium the small size of the molecules compounds with their lightness so that the value of $E_0$ is far higher than usual. While this fact by itself is not so important, it becomes extremely so if we consider some geometric facts about the helium molecules. The main thing is that they live in a three-dimensional world, so that the average separation between them does not tell us how free in its movement a molecule is. It is possible to talk about the zero point energy of a liquid only if its molecules are so close to each other and are big enough that for short periods of time some of them are effectively incased by their neighbors. When helium liquefies at $T = 4.4 \text{K}$, the separation between them is about $3.8 \text{Å}$ (the density being $\rho = 0.1216 \, \text{g} \, \text{cm}^{-3}$). Inasmuch as they have a diameter of $2.6 \text{Å}$, it is clear that already at this temperature their mean free paths are very short. At the lambda point their average separation has diminished to $3.6 \text{Å}$ (since $\rho = 0.1460 \, \text{g} \, \text{cm}^{-3}$), and it becomes evident that they are obstructing each other a lot, so much that they are almost incasing each other.

Furthermore, at any moment in time there are local fluctuations $\Delta\rho$ of either sign in the density. In positive density fluctuations the molecules are going to obstruct each other's motion even more for the duration. The incased molecules, due to Heisenberg's uncertainty principle, have to have a large wavenumber and energy. If at the lambda point the density is such that molecules are almost incasing each other, then at any higher density each molecule would certainly be incased by its neighbors, and the energetic cost to the liquid would be unaffordable. This argument may seem rough, but something like this must be precisely what is happening, since it is a fact that the density of liquid helium increases rapidly as the temperature goes down from $4.4\mathrm{K}$ to the lambda point, and then stays basically constant till absolute zero. What is special about the lambda point is that all the molecules are just on the verge of trapping and restricting each other to a very small space. It does not matter how much the pressure is increased, 25 or more atmospheres, the density cannot increase, as is experimentally observed.

Let us calculate the wavenumber for one of the temporarily incased molecules. Since approximately each one is separated from the others $3.6\mathring{\mathrm{A}}$ , and since the closest together two can get is $2.6\mathring{\mathrm{A}}$ (at this separation the potential energy grows very abruptly), then the movement available to a molecule in one direction is $3.6 - 2.6 = 1.0\mathring{\mathrm{A}}$ . From the center of the incasing box the molecule moves towards the right or the left about $\Delta x$ , so that $2\Delta x \approx 1.0\mathring{\mathrm{A}}$ , or $\Delta x \approx 0.5\mathring{\mathrm{A}}$ . Substituting this value for $\Delta x$ in Heisenberg's uncertainty principle $\Delta x\, \Delta k_x \approx \frac{1}{2}$ , we find $\Delta k_x \approx 1.0\mathring{\mathrm{A}}^{-1}$ . Obviously, for an incased molecule, $\langle k_x \rangle = 0$ . With this result and $\Delta k_x^2 = \langle k_x^2 - \langle k_x \rangle^2 \rangle$ , we immediately conclude that

$$k_{zpe} = \left( \left\langle k_x^2 + k_y^2 + k_z^2 \right\rangle \right)^{1/2} \approx 1.7\mathring{\mathrm{A}}^{-1} \quad .$$

(4)

This simple derivation based on the zero point energy effect predicts that many molecules will have a wavenumber of $1.7\mathring{\mathrm{A}}^{-1}$ , very close to the $1.9\mathring{\mathrm{A}}^{-1}$ of the experimentally observed peak.

The motion of an incased atom is, of course, invisible to the eye. Although there is a macroscopic number of atoms with the same wavenumber, they are not in the same state, because the wavevector $\boldsymbol{k}$ is pointing in a different direction for each atom. If for some reason the wavectors of the incased molecules aligned, then it would then be possible to speak of Bose-Einstein condensation at some $\boldsymbol{k}_{ZPE}$ . This effect would be observable as a macroscopic current. It goes without saying that only a few of the molecules are incased at any one moment in time, and that this is an average effect. When one fluctuation disappears, somewhere else another one appears, so that the number of incased molecules stays the same.

Suppose a pressure or temperature gradient is applied to liquid helium II, so that some of the $\boldsymbol{k}$ vectors of the incased atoms become aligned. Then a macroscopic current density $\boldsymbol{J}$ would be very quickly formed because now the vectors $\boldsymbol{k}$ are pointing on the average in the direction determined by the gradient. The kinetic energy of an atom belonging to this current cannot decrease: it is fixed by Heisenberg's uncertainty principle. Furthermore, the gradient just aligns wavevectors, so a weak gradient can quickly cause a strong current. One is reminded of G. V. Chester's dictum: "...superfluidity can be characterized by the response of the system to arbitrary weak forces..." [14]

## 3. A FINAL COMMENT AND A PROPOSAL FOR AN EXPERIMENT

The explanation for the behaviour with temperature of the peak at $Q = 1.9 \text{Å}^{-1}$ presented here is based on two physical concepts: First, that the size of helium molecules and their average separation at the lambda point are the just the right ones to *almost* make a kind of *three-dimensional* box around each molecule. Second, the uncertainty principle, which predicts that a macroscopic number of molecules are going to have a wavenumber $k = 1.7 \text{Å}^{-1}$ [11] for the size of the boxes formed, so that the most frequent momentum transfers are going to be precisely for $Q = k \approx 1.7 \text{Å}^{-1}$, very close to what is experimentally seen. This model gives a simple explanation for the observed peak in the dynamic structure factor. Other explanations, that postulate that the condensate forms at $k = 0 \text{Å}^{-1}$ but that for some complicated reason it is experimentally appearing as a peak at $1.9 \text{Å}^{-1}$, are certainly more complex, but this is not an asset in physics.

There is an experiment that can be performed to settle if the peak's formation at the lambda point corresponds to an actual increase in the number of excitations with wavenumber $k = 1.9 \text{Å}^{-1}$, like we are proposing, or if it is just an induced effect. It would consist of shining monochromatic light on liquid helium of a frequency that excites the atoms, so that some of them return to the fundamental state with the emission of a photon. The width of the spectroscopic lines produced by the decaying atoms is then measured as a function of the temperature. One would expect in general the width to diminish with temperature, but if some of the atoms at the lambda point are acquiring more kinetic energy at the cost of the rest, then the Doppler effect should begin *to broaden* the spectroscopic lines of the photons emitted by those molecules, even as the temperature continues to drop. There would be macroscopically many molecules traveling at a speed $v \approx 230\,\mathrm{m\,s^{-1}}$ .

## REFERENCES

[1] A. D. B. Woods and E. C. Svensson, Phys. Rev. Lett. **41,** 974 (1978).

[2] E. F. Talbot, H. R. Glyde, W. G. Stirling, and E. C. Svensson, Phys. Rev. **B38**, 11229 (1988).

[3] W. G. Stirling and H. R. Glyde, Phys. Rev. **B41,** 4224 (1990).

[4] K. Huang, *Statistical Mechanics*, (John Wiley & Sons, New York 1963).

[5] H. R. Glyde and A. Griffin, Phys. Rev. Lett. **65,** 1454 (1990).

[6] D. Pines, *Quantum Fluids*, ed. D.. F. Brewer (John Wiley & Sons, New York 1966), 257.

[7] A. Griffin, *Excitations in a Bose-Condensed Liquid*, (Cambridge University Press, Cambridge 1993).

[8] Y. A. Nepomnyashchy, Phys. Rev. **B46**, 6611 (1992).

[9] S. R. Shenoy, A. C. Biswas, J. Low-Temp. Phys. **28,** 191 (1977).

[10] V. I. Yukalov, Physica **100A,** 431 (1980).

[11] V. I. Yukalov Phys. Lett. **83A,** 26 (1981).

[12] G. J. Hyland, G. Rowlands and F. W. Cummings, Phys. Lett. **31A,** 465 (1970).

[13] A. Griffin, Phys. Rev. B22, 5193 (1980); G. V. Chester and L. Reatto, Phys. Rev. B22, 5199 (1980).

[14] G. V. Chester, *Proceedings of the Fifteenth Scottish Universities Summer School in Physics, 1974,* J. G. M. Armitage, I. E. Farquhar, eds., (Academic Press, London 1975), 24.